\documentclass[aps,10pt,prb,twocolumn,letterpaper,superscriptaddress]{revtex4-1}
\usepackage{amsmath}
\usepackage{dcolumn}
\usepackage{epsfig}
\usepackage{graphicx}
\usepackage{latexsym}
\usepackage{amssymb}
\usepackage{color}
\usepackage{amsfonts}
\usepackage{bm}
\usepackage{upgreek}

\newcommand{\lio}{Li$_2$IrO$_3$}
\newcommand{\nio}{Na$_2$IrO$_3$}
\newcommand{\sio}{Sr$_2$IrO$_4$}
\newcommand{\rucl}{RuCl$_3$}
\newcommand{\glio}{$\gamma$-Li$_2$IrO$_3$}
\newcommand{\blio}{$\beta$-Li$_2$IrO$_3$}
\newcommand{\um}{$\upmu$m}
\newcommand{\ums}{$\upmu$m$^2$}
\begin{document}
\title{Resonant x-ray scattering reveals possible disappearance of magnetic order under hydrostatic pressure in the Kitaev candidate \glio}

\author{Nicholas P. Breznay}
\altaffiliation{nbreznay@berkeley.edu}
\affiliation{Department of Physics, University of California, Berkeley, Berkeley CA 94720, USA}
\affiliation{Materials Science Division, Lawrence Berkeley National Laboratory, Berkeley CA 94720, USA}
\author{Alejandro Ruiz}
\affiliation{Department of Physics, University of California, Berkeley, Berkeley CA 94720, USA}
\affiliation{Materials Science Division, Lawrence Berkeley National Laboratory, Berkeley CA 94720, USA}
\author{Alex Frano}
\affiliation{Department of Physics, University of California, Berkeley, Berkeley CA 94720, USA}
\affiliation{Advanced Light Source, Lawrence Berkeley National Laboratory, Berkeley CA 94720, USA}
\author{Wenli Bi}
\affiliation{Advanced Photon Source, Argonne National Laboratory, Argonne, Illinois 60439, USA}
\affiliation{Department of Geology, University of Illinois at Urbana-Champaign, Urbana IL 61801, USA}
\author{Robert J. Birgeneau}
\affiliation{Department of Physics, University of California, Berkeley, Berkeley CA 94720, USA}
\author{Daniel Haskel}
\affiliation{Advanced Photon Source, Argonne National Laboratory, Argonne, Illinois 60439, USA}
\author{James G. Analytis}
\affiliation{Department of Physics, University of California, Berkeley, Berkeley CA 94720, USA}
\affiliation{Materials Science Division, Lawrence Berkeley National Laboratory, Berkeley CA 94720, USA}

\date{\today}

\begin{abstract}

Honeycomb iridates such as \glio{} are argued to realize Kitaev spin-anisotropic magnetic exchange, along with Heisenberg and possibly other couplings. While systems with pure Kitaev interactions are candidates to realize a quantum spin liquid ground state, in \glio{} it has been shown that the balance of magnetic interactions leads to the incommensurate spiral spin order at ambient pressure below 38~K. We study the fragility of this state in single crystals of \glio{} using resonant x-ray scattering (RXS) under applied hydrostatic pressures of up to 3.0~GPa. RXS is a direct probe of the underlying electronic order, and we observe the abrupt disappearance of the \textbf{q}=(0.57, 0, 0) spiral order at a critical pressure $P_c = 1.5$\,GPa with no accompanying change in the symmetry of the lattice. This dramatic disappearance is in stark contrast with recent studies of \blio{} that show continuous suppression of the spiral order in magnetic field; under pressure, a new and possibly nonmagnetic ground state emerges.

\end{abstract}

\maketitle

\begin{figure*}[htb!] 
	\includegraphics[width=2.0\columnwidth]{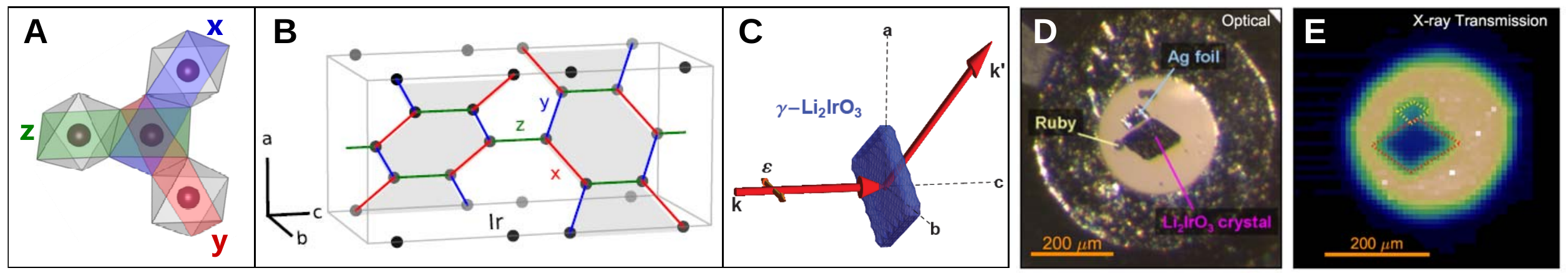}
	\caption{(A) Local geometry of edge-sharing IrO$_6$ octahedra, and (B) 3D network of Ir atoms composed of intersecting honeycomb layers (gray) in the orthorhombic crystal structure of \glio; $x$, $y$, $z$ Kitaev bonds are highlighted in both panels. (C) Laue (transmission) scattering geometry; x-rays with initial wavevector \textbf{k} and polarization $\epsilon$ scatter into \textbf{k$'$} within the \lio{} crystallographic \textit{a}-\textit{c} plane. (D) Micrograph of the loaded diamond anvil cell with polished single crystal, and Ag foil and ruby spheres for pressure calibration and \textit{in-situ} monitoring. (E) X-ray transmission image of the sample \textit{in-situ}; both \glio{} crystal and Ag foil are outlined.}
	\label{f:f1}
\end{figure*}

\textit{Introduction-}Honeycomb magnetic materials with strong spin-orbit coupling were recently proposed to realize spin-anisotropic ``Kitaev'' magnetic exchange~\cite{jackeli_mott_2009} and therefore to host a highly entangled spin-liquid ground state with fractionalized excitations~\cite{kitaev_anyons_2006, witczak_correlated_2014}. Additional interactions (such as Heisenberg or next-nearest-neighbor couplings) that compete with the Kitaev exchange leads to complex landscapes of possible spin orders~\cite{khaliullin_kitaev-heisenberg_2010, singh_relevance_2012, lee_theory_2015, kimchi_unified_2015, winter_challenges_2016}. For example, the layered compounds \nio{} and \rucl{}, composed of edge-sharing IrO$_6$ or RuCl$_6$ octahedra, show a ``zig-zag'' spin texture that onsets below 10-15~K~\cite{chaloupka_zigzag_2013, johnson_monoclinic_2015, sears_magnetic_2015}. The three-dimensional (3D) `harmonic' honeycomb $\beta$ and $\gamma$ polytypes of \lio~\cite{modic_realization_2014, takayama_hyperhoneycomb_2015} both exhibit an incommensurate spiral order~\cite{biffin_unconventional_2014, biffin_noncoplanar_2014}. These magnetic ground states derive from the balance of Kitaev (K), Heisenberg (J), and other possible couplings between spins $\vec{S}_i$ in the Hamiltonian
\begin{equation}
H = \sum_{\substack{i,j \\ \gamma \in x,y,z}} \left( {K S_i^{\gamma} S_j^{\gamma}} + J \vec{S}_i \cdot \vec{S}_j + \dots \right).
\end{equation}
Here the $\gamma = x, y, z$ Kitaev exchange directions couple spins that are perpendicular to the planes formed by adjacent IrO$_6$ octahedra, shown for \glio{} in Fig.~\ref{f:f1}AB.

Precisely tuning the relative strength of these magnetic interactions remains an outstanding challenge. Recent resonant x-ray scattering (RXS) studies of \blio{} indicate that a magnetic field both suppresses the spiral order and stabilizes a canted zig-zag spin texture~\cite{ruiz_beta_2016}, while magnetic circular dichroism studies of \blio{} under hydrostatic pressure (and also in applied fields) suggest a disappearance of the magnetic order near 2~GPa. Studies of \sio{} indicate quenching of the ferromagnetic moments at 17\,GPa~\cite{haskel_pressure_2012} and demonstrate that hydrostatic pressure can be a useful control parameter in these materials. However, to date there have been no direct probes of the spiral (or other non-ferromagnetic) order under pressure in the honeycomb iridates.

An unambiguous probe of electronic orders resolved in $Q$-space is RXS. In this technique, the energy of incident x-rays is tuned to be on resonance with an element's absorption edge and so it is sensitive to both charge and magnetic order of the valence electrons~\cite{Hannon}. At the Ir $L_{2,3}$-edge, for example, the intermediate states in the scattering process are sensitive to both the spin and orbital character of the 5d hole states. RXS is particularly useful when neutron scattering is rendered unfeasible by small sample sizes or elements with large neutron absorption cross sections like iridium. Thus, the enhanced cross section at the Ir $L_{3}$-resonance along with the large Ewald sphere afforded by 11.215~keV x-rays have made RXS the best suited scattering technique to investigate the ambient pressure magnetic order in the honeycomb iridates \glio{}~\cite{biffin_noncoplanar_2014}, \blio{}~\cite{biffin_unconventional_2014}, and \nio{}~\cite{Liu_lio_RXS}. Conducting RXS at the Ir $L_{3}$-edge under applied pressure is possible in reflection and transmission modes~\cite{haskel_sio}, but both modes are constrained by the geometry of a high pressure apparatus and resulting limitations on the sample geometry and available reciprocal space. Hence magnetic RXS at non-ambient pressures represents an acute experimental challenge.

In this work, we modify the magnetic ground state in the 3D honeycomb iridate \glio{} by applying hydrostatic pressure and observe the disappearance of the incommensurate spiral order using RXS. In contrast to recent studies that use symmetry-breaking magnetic fields~\cite{takayama_hyperhoneycomb_2015, ruiz_beta_2016}, here we observe disappearance of the spiral order with no observed discontinuity in the lattice structure to the highest pressures measured. The abrupt disappearance of the spiral Bragg peak at a critical pressure $P_c = 1.5$\,GPa signals the transition to a distinct electronic ground state.

\textit{Experimental-}We track the disappearance of the spiral magnetic order in \glio{} using RXS on oriented single crystal samples. The transmission scattering scheme, shown in Fig.~\ref{f:f1}C, allows access to the ($H$ 0 0) direction in specular diffraction using horizontally ($\mathbf{\sigma}$) polarized photons in a vertical geometry and with no polarization analysis of the scattered beam. The magnetic scattering intensity, proportional to $|\sum_{i}e^{i\mathbf{Q}\cdot \mathbf{r}_i}(\mathbf{\sigma} \times \mathbf{\pi}_{out})\cdot \mathbf{m}_i|^2$~\cite{Hannon}, where $\mathbf{m}_i$ is the magnetic moment at site $r_i$, projects out the component of $\mathbf{m}_i$ parallel to \textit{k$'$} (in the \textit{a}-\textit{c} scattering plane).

Crystals of \glio{} were grown as described previously~\cite{modic_realization_2014, ruiz_beta_2016}. Figure~\ref{f:f1}B shows the intersecting Ir honeycomb layers in one unit cell of the orthorhombic Cccm crystal structure. For the transmission (Laue) scattering geometry, we polished single crystals to 20-30\,\um\ thick; the absorption length of x-rays near the Ir L$_{2,3}$ edges is $\approx10$\,\um.

X-ray scattering studies under pressure (up to 5\,GPa) and at temperatures between 5-300\,K were performed at beamline 4-ID-D of the Advanced Photon Source at Argonne National Lab. Merrill-Bassett type diamond anvil cells (DAC) with 800\,\um\ culets were used with stainless steel gaskets of 250 (150)\,\um\ initial (pre-indented) thicknesses, with 400\,\um\ sample chamber holes~\cite{Feng_High_2010}. The gaskets were loaded with \glio{} single crystals (cross-sectional area $150 \times 100$ \ums, polished thickness $25$\,\um), along with several ruby balls and $40 \times 40$\,\ums\ pieces of 12\,\um\ thick Ag foil for ambient- and low-temperature pressure calibration~\cite{Holzapfel_Equations}. The pressure medium was a 4:1 methanol:ethanol mixture. After initial loading, the pressure at ambient temperature was monitored using a custom-built optical spectrometer and a Raman system to measure the ruby $R$1 fluorescence peak; the target pressure upon loading was $\sim$0.1\,GPa. The pressure at low temperature was determined in situ using Ag powder peaks and the isothermal bulk modulus of Ag at 5\,K ($K_{\textrm{Ag}}$=110.85\,GPa, $K'_{\textrm{Ag}}$=6.0\,GPa)~\cite{Holzapfel_Equations}. We estimate a systematic uncertainty of $\pm$0.1\,GPa in the pressures quoted below by comparing the estimated pressure from (1 1 1), (2 0 0), and (2 2 0) Ag powder peaks, and from repeated pressure measurements before and after scans.

Two samples were cooled, aligned in the diffractometer, and studied; all measurements reported here were performed at the cryostat base temperature of 4.5$\pm$0.5\,K. Pressure was changed \textit{in situ} using a helium membrane. Cell layout and sample status were checked after loading (see optical image in Fig.~\ref{f:f1}D, showing diamond-shaped \glio{} crystal, Ag foil, and ruby spheres in the DAC gasket hole) and monitored in-situ using both x-ray transmission maps (Fig.~\ref{f:f1}E, obtained using a slit-defined $30\times30 \ \mu$m$^2$ beam). To track the absolute magnetic Bragg peak intensities with pressure, peak areas were normalized to the integrated (4 0 0) rocking curve intensities. The mosaic full-width at half-maximum is 0.05-0.10$^{\circ}$ for sample 1 between 0-3\,GPa, and 0.01$^{\circ}$ for sample 2 at 2.0\,GPa.

\begin{figure}[htb]
\includegraphics[width=1.0\columnwidth]{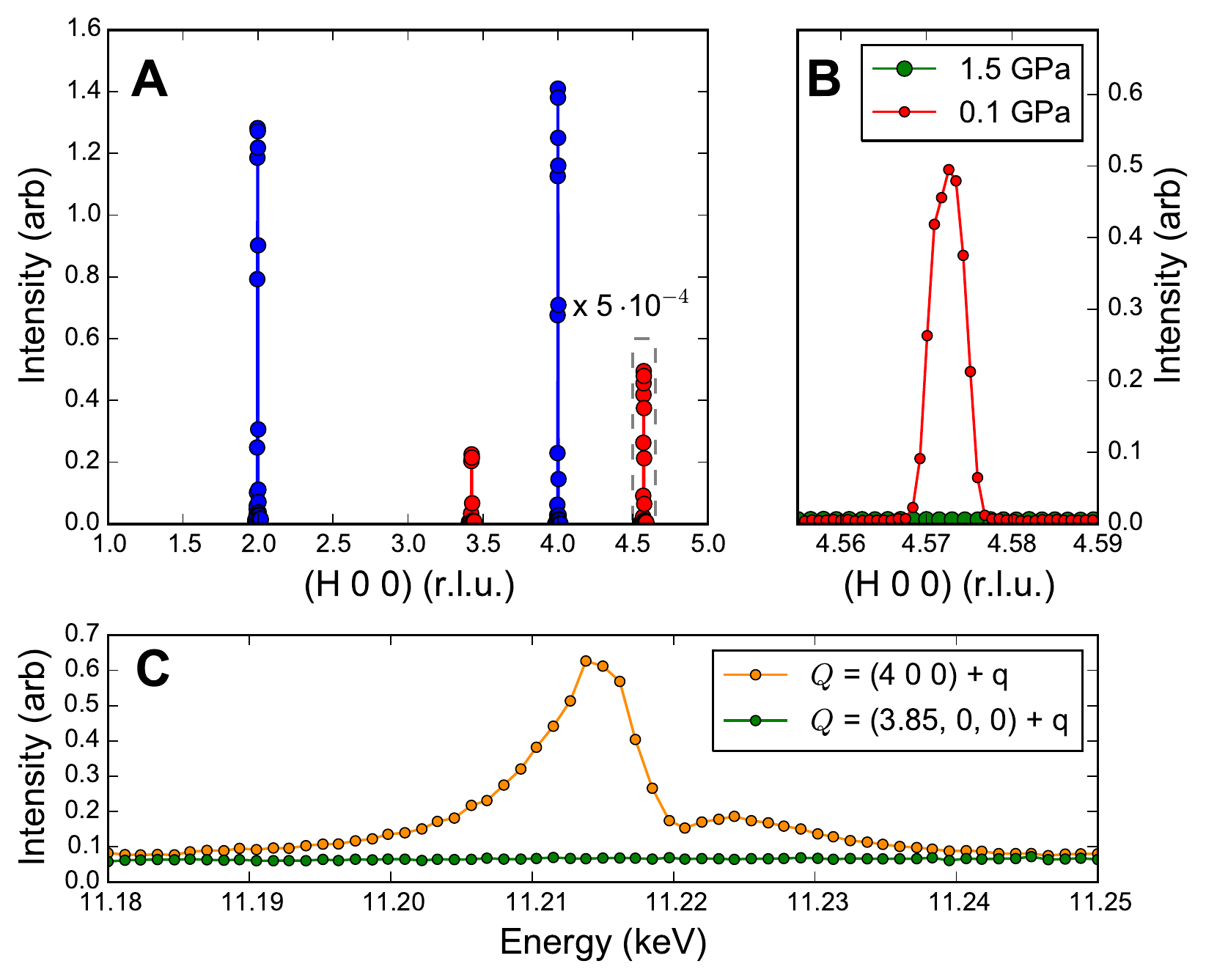}
\caption{(A) Magnetic (red) and structural (blue) Bragg peaks along the (H 0 0) direction measured at 4.5\,K; the (4 0 0) peak has been scaled by a factor of 5$\times 10^{-4}$. (B) Zoom-in of the dashed region in (A) showing scans near (4 0 0) + \textbf{q} = (4.57 0 0). The magnetic Bragg peak visible at 0.1\,GPa (red) disappears abruptly above 1.5\,GPa (green). (C) Energy scans at fixed \textbf{Q} = (4.57 0 0) (orange) and away from the magetic peak \textbf{Q} = (4.42 0 0) showing a featureless background.}
\label{f:f2}
\end{figure}

\textit{Results-}Based on the restricted scattering geometry imposed by the DAC, we focused our study on (H 0 0) peaks in reciprocal space. In \glio, selection rules forbid the (1 0 0) and (3 0 0) lattice peaks, while the structure factor for the (2 0 0) order is strongly suppressed. Figure~\ref{f:f2}A shows a reciprocal space map of the (H 0 0) axis, where an intense structural (4 0 0) Bragg peak and weaker (2 0 0) peak are both observed. (In the \blio{} polytype the (2 0 0) peak is forbidden.) Non-structural peaks located at (4$\pm$0.57 0 0) correspond to the incommensurate spiral magnetic order~\cite{biffin_noncoplanar_2014, biffin_unconventional_2014}, which at ambient pressure onsets at $T_{\textrm{sp}}=38$\,K. The electronic nature of these peaks is confirmed by fixed-$Q$ energy scans (Fig.~\ref{f:f2}C) showing a strong enhancement of the diffracted intensity near the 11.215\,keV Ir L$_3$ resonance, in contrast to the weak background observed away from the spiral order peak at (3.85 0 0) + \textbf{q} that shows an increase of $\sim$10\% above the Ir edge due to fluorescence.

\begin{figure}[htb]
\includegraphics[width=1.0\columnwidth]{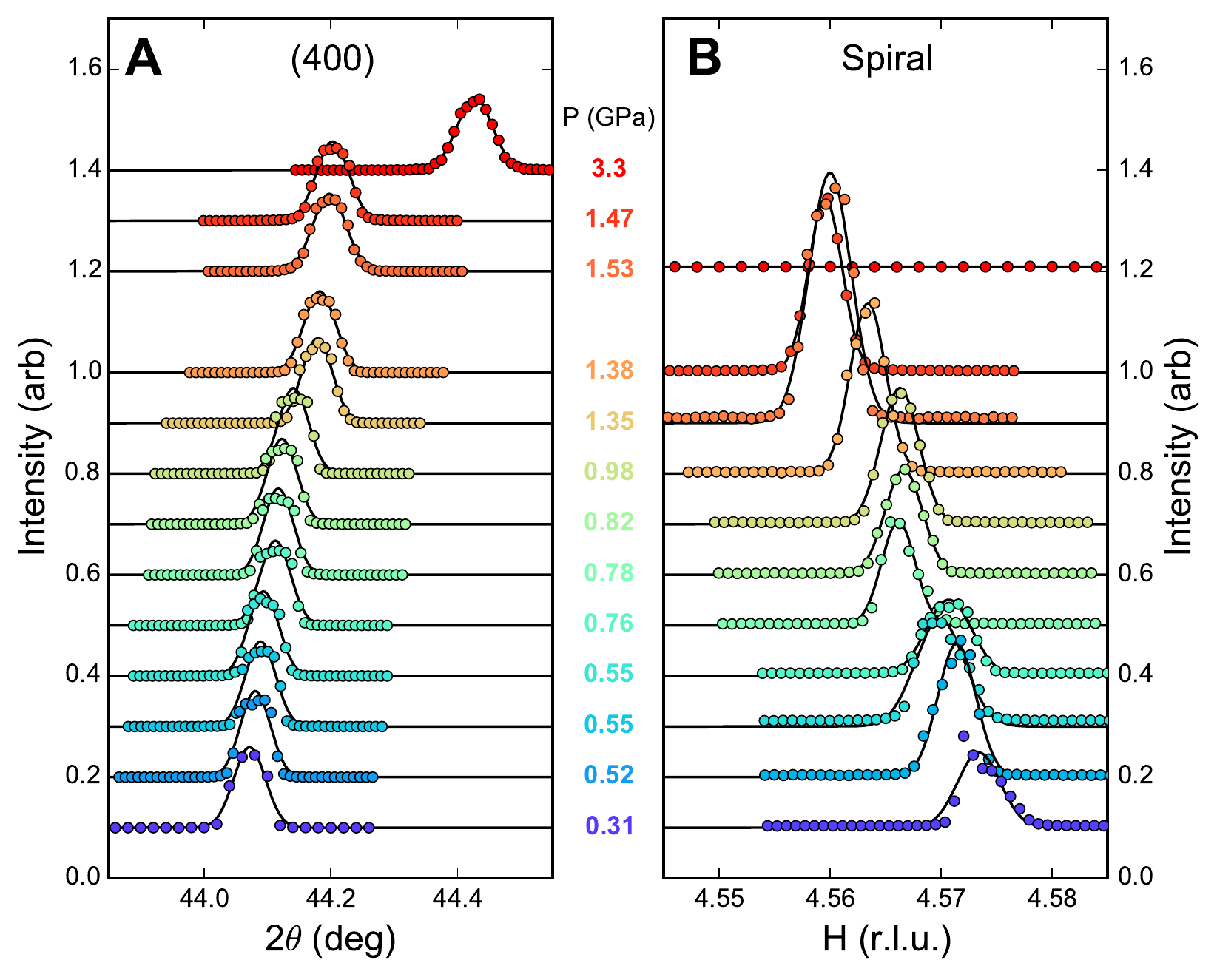}
\caption{(A) (400) Bragg peak intensity versus 2$\theta$ and (B) evolution of spiral order (4.57 0 0) peaks with applied pressures from 0.3 to 3.0\,GPa as labeled for both plots. Solid lines are guides to the eye.}
\label{f:f3}
\end{figure}

\begin{figure}[tb]
\includegraphics[width=1.0\columnwidth]{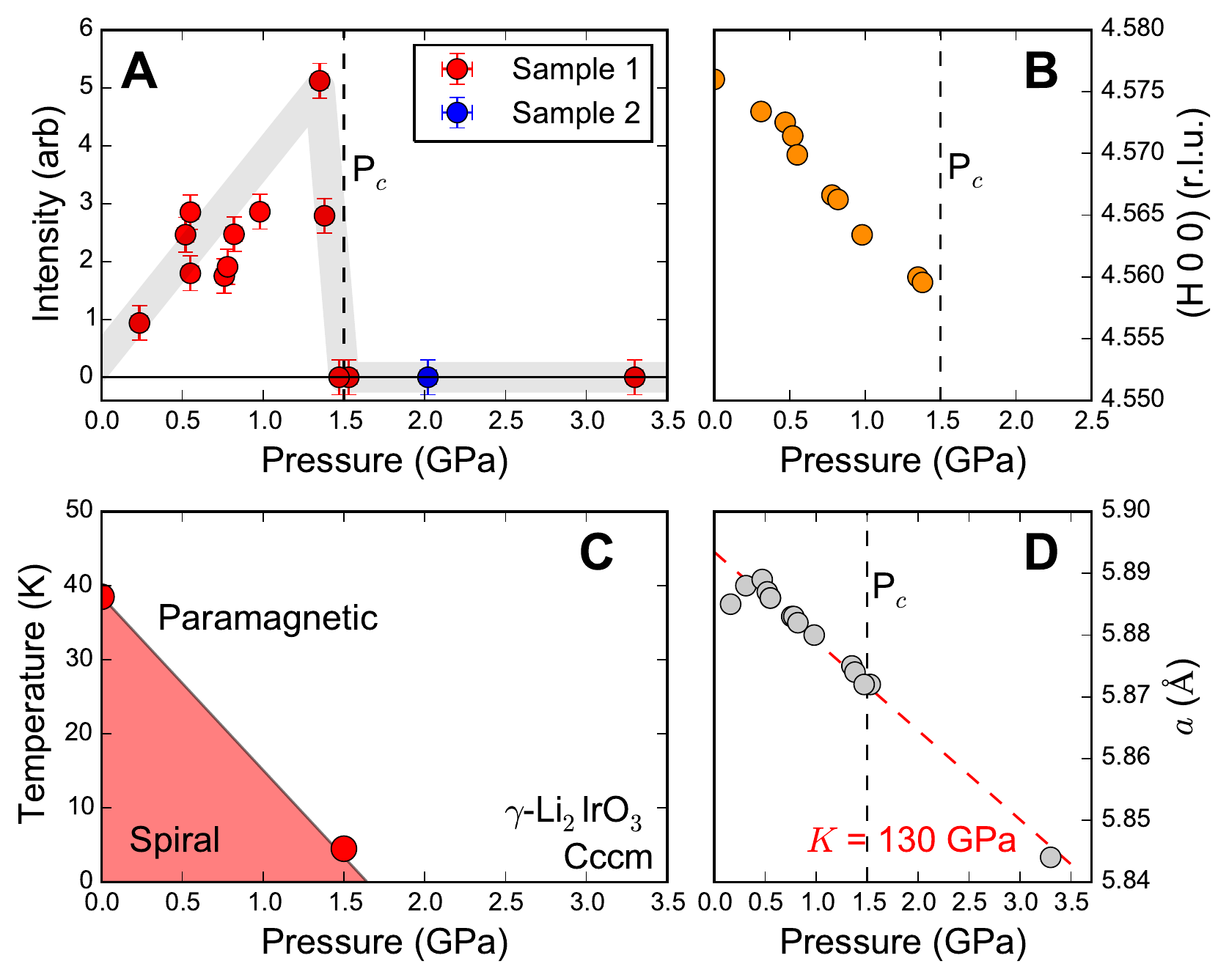}
\caption{(A) Magnetic Bragg peak intensity versus applied pressure for two samples; the intensity for sample 1 disappears abruptly at $P_c$=1.5\,GPa; no magnetic is observed at 2.0\,GPa for sample 2. (B) Decrease in the spiral order wavevector \textbf{Q}=(H 0 0) with applied pressure; the wavevector is not close to a commensurate value at $P_c$. (C) Pressure-temperature magnetic phase diagram for \glio; no discontinuous change in the lattice is observed to 3.0\,GPa. (D) Decrease in \textit{a} with applied hydrostatic pressure, extracted from the (4 0 0) structural Bragg peak; assuming a relative change in volume that is isotropic, the bulk modulus $K = 130$\,GPa.}
\label{f:f4}
\end{figure}

Figure~\ref{f:f3} shows the pressure dependence of the (4 0 0) Bragg peak (normalized), as well as the spiral order peak. Note that the shift in the 3.3\,GPa (4 0 0) scan is a consequence of a large pressure increment; the \textit{a}-axis lattice parameter evolves linearly with pressure over the entire range studied. Neither the (4 0 0) Bragg peak nor \textbf{q} peak widths change appreciably with increasing pressure, indicating that both the crystal quality and long range nature of the spiral order remain constant. Figure~\ref{f:f4}D shows the evolution of the \textit{a}-axis lattice parameter with pressure, along with a linear fit yielding $d\textit{a}/dP = -0.015 \textrm{\AA}/\textrm{GPa}$. Assuming an isotropic fractional change in the unit cell dimensions, the T~=~4.5\,K bulk modulus $K = -\frac{1}{3}\textit{a}(\frac{d\textit{a}}{dP})^{-1} = 130\pm20$\,GPa. Recent electronic structure calculations for \blio{} indicate an anisotropic compressibility~\cite{Kim_Revealing} with \textit{a} more compressible than \textit{c}; this would increase $K$ relative to the above estimate and be quite comparable to the 150-250\,GPa observed in \sio{} and Sr$_3$Ir$_2$O$_7$ compounds~\cite{haskel_sio, zhao_pressure_2014}, and recent studies of \blio{} powders~\cite{haskel_beta}.

Under an applied pressure of $P_c = 1.5$\,GPa, the spiral order peaks are abruptly extinguished as shown in Fig.~\ref{f:f2}B for peaks at (4 0 0) + \textbf{q}. We scanned the entire accessible range of 2$<$H$<$6 r.l.u. and found no evidence for the incommensurate peaks anywhere in this region. Aside from contraction of the unit cell (shown in Fig.~\ref{f:f4}D), no structural changes were observed; the symmetry of the lattice appears intact throughout this pressure range. Figure~\ref{f:f4}A shows the integrated peak intensity for \textbf{q} as a function of applied pressure, measured at 4.5\,K incrementally beginning at 0.5\,GPa. To within the uncertainty associated with consistent re-alignment of the sample after changing pressure, the peak intensities gradually increase with pressure until abruptly disappearing at $P_c$. An (otherwise identical) additional sample also showed no sign of the spiral order at the after cool-down pressure of 2.0\,GPa. The incommensurate wavevector \textbf{q}, normalized to the change in lattice constant, decreases with increasing pressure as shown in Fig.~\ref{f:f4}B; \textbf{q} decreases by 0.3\% before the spiral order disappears at $P_c$. The linear and continuous decrease observed in \textbf{q}, ending at an apparently irrational fraction, precludes an incommensurate-to-commensurate transition.

\textit{Discussion-}Figure~\ref{f:f4}C presents a schematic pressure-temperature phase diagram for \glio{}. Ongoing studies of this material indicate paramagnetic behavior with rapidly emerging magnetic anisotropy favoring the \textit{b} (easy) axis direction~\cite{modic_realization_2014, ruiz_betagamma_2017} at ambient pressure. With increasing pressure and as T$\rightarrow$0, $P_c$ likely continues to represent a sharp phase boundary between the spiral magnetic order and the as-yet undetermined high-pressure electronic phase. As there is no change in the lattice symmetry, and no symmetry-breaking field being used to perturb the material, it is unclear what ordered state may exist beyond $P_c$, if any. Under an applied magnetic field, a canted zig-zag order appears to be the next competitive magnetic ground state at ambient pressure~\cite{ruiz_beta_2016}. However the transition observed under applied fields appears to be continuous, in contrast to the transition at $P_c$, indicating that a different ground state emerges in this case.

The effect of pressure on the crystal structure and associated Kitaev, Heisenberg, and other exchange couplings, was recently studied theoretically for \blio{}~\cite{Kim_Revealing}. In this closely related polytype, the Kitaev exchange coupling was predicted to decrease with increasing pressure above ambient, disappearing at 5-10\,GPa. If similar evolution of the magnetic interactions appears in \glio{}, our results strongly suggest that the spiral magnetic order is stabilized by Kitaev exchange and suppressed as this mechanism weakens. Precise experimental description of structural distortion with increasing pressure in both \lio{} polytypes will allow for a quantitative analysis of how the change in structure serves to push this material closer to or farther from the pure Kitaev limit.

The scale of $P_c$ is modest compared to the 17\,GPa required to suppress weak ferromagnetism in \sio{}~\cite{haskel_pressure_2012}. The pressure
roughly corresponds to an energy density of 9\,meV/\AA$^3$, or 0.08\,eV/Ir; while less than both the spin-orbital energy $\lambda_{\textrm{SO}} \sim 0.2-0.5$\,eV as well as electronic interaction U$\sim 0.5$\,eV that have been reported in 5d iridate materials~\cite{lagunamarco_orbital_2010, haskel_pressure_2012, clancy_spin-orbit_2012}, it is beyond the $\sim 1$\,meV scale that was proposed to separate \blio{} from a 3D spin liquid state~\cite{katukuri_vicinity_2016}. Mixing of the (nominally filled) J$ = 3/2$ manifold of states could serve to disrupt the J$_\textrm{eff} = 1/2$ doublet that is crucial to the stability of unconventional magnetic orders in these materials. Such a picture could be investigated quantitatively with high-pressure studies of x-ray absorption (XAS) and magnetic circular dichroism (XMCD) spectroscopies that provide a quantitative probe of the spin and orbital components of the local magnetic moments. 

\textit{Summary-}By conducting resonant x-ray scattering studies at the Ir $L_3$ edge, we are able to observe the disappearance of the spiral magnetic order in \glio{} at an applied pressure of 1.5\,GPa. This observation provides strong evidence for tunability of the Kitaev, Heisenberg, and other magnetic exchange couplings with applied pressure. Future resonant diffraction studies will be able to incisively address the possibility of complete disappearance of long range magnetic order in the high pressure ground state of this Kitaev candidate material.

\begin{acknowledgments}
The authors thank Z. Islam, Y. Choi, and the support staff at APS Sector 4 for their assistance, and P.~Moll and J.~Reuteler at Scope-M ETH for additional sample preparation. This work was supported by the U.S. Department of Energy, Office of Science, Basic Energy Sciences under Award No. {DE-SC0014039}. This research used resources of the Advanced Photon Source, a U.S. Department of Energy (DOE) Office of Science User Facility operated for the DOE Office of Science by Argonne National Laboratory under Contract No. {DE-AC02-06CH11357}. NPB was supported by the Gordon and Betty Moore Foundation's EPiQS Initiative through Grant GBMF4374. AF acknowledges support from the University of California President's Postdoctoral Fellowship Program. WB acknowledges partial support from COMPRES, the Consortium for Materials Properties Research in Earth Sciences under NSF Cooperative Agreement EAR 1606856. 
\end{acknowledgments}


\end{document}